\newif\ifpdflatex    
\pdflatextrue           

\documentclass[fleqn,usenatbib,useAMS,twocolumn]{aastex}
\usepackage{amsmath,esint}
\usepackage{url,aasmacros}
\usepackage{natbib}
\usepackage{soul, CJK}
\usepackage{multirow}

\usepackage{tablefootnote}

\usepackage{bm}
\usepackage{xspace}
\usepackage{color}
\usepackage{enumitem}
\usepackage{booktabs}

\usepackage{graphicx}	
\usepackage{amsmath}	
\usepackage{amssymb}	
\usepackage{bm}		
\usepackage{url}
\usepackage{amsbsy}
\usepackage{xspace}
\usepackage{hyperref}
\usepackage{longtable}
\usepackage[colorinlistoftodos]{todonotes}
\usepackage[flushleft]{threeparttable}

\usepackage[T1]{fontenc}
\usepackage{ae,aecompl}

\def\lesssim{\mathrel{\hbox{\rlap{\hbox{\lower5pt\hbox{$\sim$}}}\hbox{$<$}}}}
\def\gtrsim{\mathrel{\hbox{\rlap{\hbox{\lower5pt\hbox{$\sim$}}}\hbox{$>$}}}}

\usepackage{upgreek}                                                  
\usepackage{xspace}                                                               
\def\apjl{ApJ}%
\def\apjs{ApJS}%
\def\aap{A\&A}%
\shorttitle{SN\,2021fcg/ ZTF\,21aaoryiz}
\shortauthors{Karambelkar et al.}

\begin{document}
\title{Faintest of them all : ZTF\,21aaoryiz/SN\,2021fcg $-$ Discovery of an extremely low luminosity Type Iax supernova}
\author[0000-0003-2758-159X]{Viraj R. Karambelkar}
\email{viraj@astro.caltech.edu}
\affiliation{Cahill Center for Astrophysics, California Institute of Technology, Pasadena, CA 91125, USA}

\author{Mansi M. Kasliwal}
\affiliation{Cahill Center for Astrophysics, California Institute of Technology, Pasadena, CA 91125, USA}

\author{Kate Maguire}
\affiliation{School of Physics, Trinity College Dublin, the University of Dublin, College Green, Dublin}

\author{Shreya G. Anand}
\affiliation{Cahill Center for Astrophysics, California Institute of Technology, Pasadena, CA 91125, USA}

\author{Igor Andreoni}
\affiliation{Cahill Center for Astrophysics, California Institute of Technology, Pasadena, CA 91125, USA}

\author{Kishalay De}
\affiliation{Cahill Center for Astrophysics, California Institute of Technology, Pasadena, CA 91125, USA}

\author{Andrew Drake}
\affiliation{Cahill Center for Astrophysics, California Institute of Technology, Pasadena, CA 91125, USA}

\author[0000-0001-5060-8733]{Dmitry A. Duev}
\affiliation{Division of Physics, Mathematics, and Astronomy, California Institute of Technology, Pasadena, CA 91125, USA}

\author{Matthew J. Graham}
\affiliation{Cahill Center for Astrophysics, California Institute of Technology, Pasadena, CA 91125, USA}

\author[0000-0002-7252-3877]{Erik C. Kool} 
\affiliation{The Oskar Klein Centre, Department of Astronomy, Stockholm University, AlbaNova, SE-10691, Stockholm, Sweden}

\author[0000-0003-2451-5482]{Russ R. Laher}
\affiliation{IPAC, California Institute of Technology, 1200 E. California Blvd, Pasadena, CA 91125, USA}

\author{Mark R. Magee}
\affiliation{School of Physics, Trinity College Dublin, the University of Dublin, College Green, Dublin}

\author[0000-0003-2242-0244]{Ashish~A.~Mahabal}
\affiliation{Division of Physics, Mathematics and Astronomy, California Institute of Technology, Pasadena, CA 91125, USA}
\affiliation{Center for Data Driven Discovery, California Institute of Technology, Pasadena, CA 91125, USA}

\author[0000-0002-7226-0659]{Michael S. Medford}
\affiliation{Department of Astronomy, University of California, Berkeley, Berkeley, CA 94720}
\affiliation{Lawrence Berkeley National Laboratory, 1 Cyclotron Rd., Berkeley, CA 94720}

\author[0000-0001-8472-1996]{Daniel Perley}
\affiliation{Astrophysics Research Institute, Liverpool John Moores Univesity, 146 Brownlow Hill, Liverpool, L3 5RF, UK}

\author[0000-0002-8121-2560]{Mickael Rigault}
\affiliation{Univ Lyon, Univ Claude Bernard Lyon 1, CNRS, IP2I Lyon / IN2P3, IMR 5822, F-69622, Villeurbanne, France}

\author[0000-0001-7648-4142]{Ben Rusholme}
\affiliation{IPAC, California Institute of Technology, 1200 E. California Blvd, Pasadena, CA 91125, USA}
             
\author[0000-0001-6797-1889]{Steve Schulze}
\affiliation{The Oskar Klein Centre, Department of Physics, Stockholm University, AlbaNova, SE-10691, Stockholm, Sweden}

\author{Yashvi Sharma}
\affiliation{Cahill Center for Astrophysics, California Institute of Technology, Pasadena, CA 91125, USA}

\author[0000-0003-1546-6615]{Jesper Sollerman}
\affiliation{The Oskar Klein Centre, Department of Astronomy, Stockholm University, AlbaNova, SE-10691 Stockholm, Sweden}

\author{Anastasios Tzanidakis}
\affiliation{Cahill Center for Astrophysics, California Institute of Technology, Pasadena, CA 91125, USA}

\author{Richard Walters}
\affiliation{Caltech Optical Observatories, Pasadena, CA 91125, USA}

\author[0000-0001-6747-8509]{Yuhan Yao}
\affiliation{Cahill Center for Astrophysics, California Institute of Technology, Pasadena, CA 91125, USA}




\date{Last up-dated \today; in original form \today}


%

\begin{abstract}
We present the discovery of ZTF\,21aaoryiz/SN\,2021fcg -- an extremely low-luminosity Type Iax supernova. SN\,2021fcg was discovered by the Zwicky Transient Facility in the star-forming galaxy IC0512 at a distance of $\approx$ 27 Mpc. It reached a peak absolute magnitude of $M_{r} =$  $-12.66\pm0.20$ mag, making it the least luminous thermonuclear supernova discovered to date. The E(B-V) contribution from the underlying host galaxy is unconstrained. However, even if it were as large as 0.5 mag, the peak absolute magnitude would be $M_{r} = -13.78\pm0.20$ mag -- still consistent with being the lowest luminosity SN. Optical spectra of SN\,2021fcg taken at 37 and 65 days post maximum show strong [Ca~II], Ca~II and Na~I D emission and several weak [Fe~II] emission lines. The [Ca~II] emission in the two spectra has extremely low velocities of $\approx 1300$ and 1000 km~s$^{-1}$ respectively. The spectra very closely resemble those of the very low luminosity Type Iax supernovae SN~2008ha, SN~2010ae and SN~2019gsc taken at similar phases. The peak bolometric luminosity of SN\,2021fcg is $\approx$ $2.5^{+1.5}_{-0.3}\times10^{40}$ erg s$^{-1}$ which is a factor of three lower than that for SN~2008ha. The bolometric lightcurve of SN\,2021fcg is consistent with a very low ejected nickel mass (M$_{\rm{Ni}} \approx 0.8^{+0.4}_{-0.5}\times10^{-3}$ M$_{\odot}$). The low luminosity and nickel mass of SN\,2021fcg pose a challenge to the picture that low luminosity SNe Iax originate from deflagrations of near M$_{\rm{ch}}$ hybrid carbon-oxygen-neon white dwarfs. Instead, the merger of a carbon-oxygen and oxygen-neon white dwarf is a promising model to explain SN\,2021fcg.
\\
\\
\end{abstract}


\section{Introduction}

Type Iax supernovae (SNe) are a peculiar subclass of Type Ia SNe \citep{Foley2013}. These events are named after the prototypical SN\,2002cx \citep{Li2003} and are characterised by slower expansion speeds (2000--8000 km s$^{-1}$) and a diverse range of luminosities compared to normal SNe Ia \citep{Jha2017}. The luminosities of SNe Iax vary from $M_{r}\approx -19$ at the bright end (SN\,2008A, \citealt{McCully2014}) to $M_{r}\approx-14$ at the faint end (SN\,2008ha, \citealt{Valenti2009,Foley2009}). SNe Iax account for $\sim 31 \%$ of the total SN Ia rate \citep{Foley2013} and are believed to be associated with thermonuclear explosions of white dwarfs \citep{Jha2017}. However, their exact progenitors and explosion mechanisms still remain unknown.

Four SNe Iax have been discovered with very low luminosities ($M_{\rm{V}} \approx -14$) and explosion energies -- SN\,2008ha ($M_{\rm{V}} = -14.2$, \citealt{Valenti2009,Foley2009}), SN\,2010ae ($-13.8 > M_{V} > -15.3$,  \citealt{Stritzinger2014}), SN\,2019gsc (M$_{r} = -13.9$, \citealt{Srivastav2020,Tomasella2020}) and SN\,2019ttf (M$_{r} \approx -14$, \citealt{De2020a}), although only the first three have been studied extensively. These SNe have low ejected nickel masses ($\sim 10^{-3}$ M$_{\rm{\odot}}$) and a faster evolution than their brighter counterparts. Several explosion mechanisms have been proposed to account for the low luminosities and nickel masses of faint SNe Iax -- partial deflagration of a hybrid CONe white dwarf \citep{Kromer2015}, merger of a CO and ONe white dwarf \citep{Kashyap2018}, a helium-nova \citep{McCully2014}, an ultra-stripped electron-capture SN \citep{Pumo2009}, a ``fallback" massive star SN \citep{Moriya2010}. However, the small sample of these events makes it difficult to distinguish between these models. 

In this Letter, we present the discovery of ZTF\,21aaoryiz or SN 2021fcg -- the least luminous member of the SN Iax class. SN\,2021fcg has peak $M_{r} =$ $-12.66\pm0.20$ and is the lowest luminosity thermonuclear SN discovered to date. Here, we present optical photometric and spectroscopic follow-up of this transient. In Section \ref{sec:basic}, we describe the discovery and details of our follow-up observations. In Section \ref{sec:lc} we analyse the light curve to derive ejecta masses. In Section \ref{sec:spectroscopy}, we present the spectroscopic evolution of this SN. In Section \ref{sec:discussion}, we discuss this SN in the context of different formation scenarios. We conclude with a summary of our results in Section \ref{sec:conclusions}.

\section{Discovery and Follow-up Observations}
\label{sec:basic}
\subsection{Discovery}
\begin{figure*}[hbt]
    \centering
    \includegraphics[width=\textwidth]{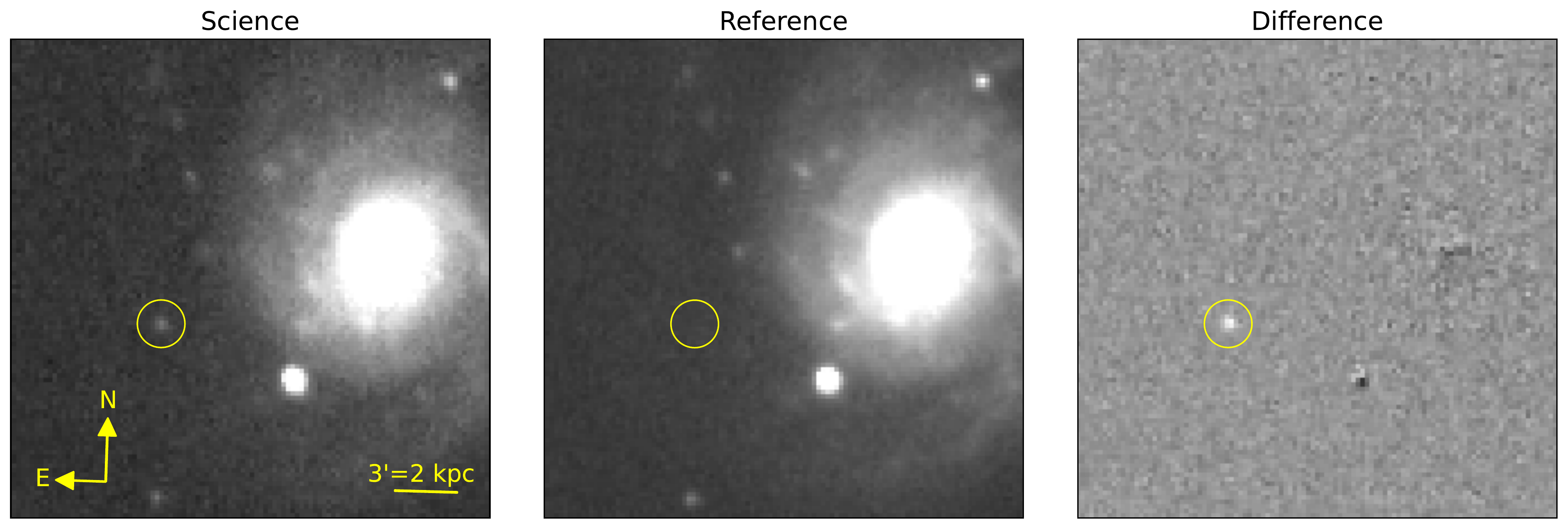}
    \caption{ZTF \emph{r-}band science, reference and difference discovery images of SN\,2021fcg. The position of the supernova is marked with a yellow circle. The science image was taken on MJD 59281.22.}
    \label{fig:disc}
\end{figure*}

SN\,2021fcg was discovered by the Zwicky Transient Facility (ZTF, \citealp{Bellm2019,Graham2019,DeKany2020}), which runs on the Palomar 48-inch (P48) Oschin Schmidt telescope. The first real-time alert \citep{Patterson2019} was generated on 20210308.22 UT (MJD 59281.22) at J2000 coordinates of $\alpha = 09^{\rm{h}}04^{\rm{m}}32.37^{\rm{s}}$, $\delta = + 85^{\rm{d}}29^{\rm{m}}48.44^{\rm{s}}$. The transient was automatically tagged as a supernova candidate by a machine-learning-based Alert-Classifying Artificial Intelligence program (ACAI; Duev et al. in prep). It was later flagged by the Census of Local Universe program (CLU, see \citealt{De2020a} for details) that identifies transients associated with nearby ($<200$ Mpc) galaxies on the Fritz portal \citep{skyportal2019,duev2019real,Kasliwal2019}. Figure \ref{fig:disc} shows the ZTF discovery image of this transient.

\subsection{Host Galaxy and Extinction}
SN\,2021fcg is located on the outskirts of the star forming spiral galaxy IC0512 at a physical separation of $\approx 7.5$ kpc (angular separation $\approx$ 53\arcsec) from the nucleus (Figure \ref{fig:disc}). The host galaxy has a redshift of $\rm{z} = 0.005384$ and a heliocentric velocity of 1614$\pm$10 km s$^{-1}$ \citep{Kourkchi2017}. Correcting for the Virgo Infall, Shapley cluster and Great Attractor  \citep{Mould2000} gives a distance modulus $\mu = 32.14 \pm 0.15$ mag (we use H$_{0} = 73$ km s$^{-1}$ Mpc$^{-1}$ in this paper) \footnote{\citet{Theureau07} report a Tully-Fisher distance modulus of $31.82\pm0.41$ mag using H$_{0} = 57$ km s$^{-1}$. Adopting H$_{0} = 73$ km s$^{-1}$ Mpc$^{-1}$ will reduce this distance modulus, making SN 2021fcg even lower luminosity than reported here}. 

The Galactic extinction along the line of sight to this galaxy is $A_{V} = 0.208$ mag \citep{Schlafly11}, for a standard reddening law with $R_{V} = 3.1$. We cannot estimate extinction due to the host galaxy accurately as our spectra do not show any absorption lines from the host. However, the supernova is located in the outskirts of the host galaxy where the host extinction is likely low (Figure \ref{fig:disc}). We thus adopt $E(B-V)_{\rm{tot}}$ = $E(B-V)_{\rm{MW}}$ =  0.068 mag. We discuss the implications of host extinction on our absolute magnitude estimates in Section \ref{sec:lc}.

\subsection{Follow-up observations}
The field containing SN\,2021fcg was observed on several epochs by the ZTF camera on P48 in the \emph{g}, \emph{r} and \emph{i} bands. The images were processed by the ZTF Data System Pipelines \citep{Masci2019} which perform image subtraction based on the ZOGY algorithm \citep{Zackay2016}.  We performed forced point spread function photometry at the location of the transient in all subtracted ZTF images. We obtained additional \emph{g}, \emph{r}, \emph{i} bands photometric observations with the Spectral Energy Distribution Machine (SEDM, \citealt{Blagorodnova2018,Rigault2019}) mounted on the 60 inch telescope at Palomar (P60) on MJD 59306 and 59318. We also observed the field in \emph{r}-band with the Alhambra Faint Object Spectrograph and Camera (ALFOSC) at the 2.56m Nordic Optical Telescope (Spain) on MJD 59367. We reduced the data with the PyNOT\footnote{\href{https://github.com/jkrogager/PyNOT}{ https://github.com/jkrogager/PyNOT}} pipeline. Finally, the field was also observed by the ATLAS survey \citep{Tonry2018,Smith2020}. We used the ATLAS forced photometry service  \footnote{https://fallingstar.com} to query all photometric measurements at the location of SN\,2021fcg. We combined measurements from same-day observations by taking a weighted mean of the flux using the inverse of the square of the flux uncertainties as weights.  
All photometric measurements (3-$\sigma$ detections and 5-$\sigma$ upper limits) are listed in Table \ref{tab:phot}.

Our spectroscopic followup comprises of two optical spectra obtained with the Low Resolution Imaging Spectrograph (LRIS, \citealt{Oke95}) on the Keck I 10 m telescope. The spectra were obtained on MJD 59318 and 59344 corresponding to +37 days and +63 days after the \emph{r}-band maximum of SN\,2021fcg. The spectra were reduced using the IDL-based tool \texttt{lpipe} \citep{Perley2019}.  

\begingroup
\renewcommand{\tabcolsep}{3pt}
\begin{table*}
\begin{center}

\caption{Photometric measurements of SN\,2021fcg ($>3\sigma$ detections and 5$\sigma$ limits, data behind Figure \ref{fig:lc})}
\label{tab:phot}
\begin{tabular}{cccccccc}
\hline
\hline
{MJD} & {Phase$^{a}$} & {\emph{g}} & {\emph{r}} & {\emph{i}} & {\emph{c}} & {\emph{o}} & {Instrument}  \\ 
\hline
59256.25 & $-27.7$ & $>20.77$ & $>20.90$ & -- & -- & -- & ZTF \\
59265.28 & $-18.7$ & $>20.59$ & $>20.62$ & -- & -- & &ZTF \\
59270.45 & $-13.5$ & -- & -- & -- & -- & $>19.27$ & ATLAS \\
59275.16 & $-08.8$ & $20.58 \pm 0.23$ & $>19.96$ & -- &-- & -- & ZTF \\
59276.61 & $-07.4$ & -- & -- & -- & -- & $19.63\pm0.24$ & ATLAS \\
59278.25 & $-05.7$ & $20.25 \pm 0.22$ & -- & -- & --  & -- & ZTF \\
59278.49 & $-05.5$ & -- & -- & -- & -- & $19.79\pm0.31$ & ATLAS \\
59281.22 & $-02.8$ & $20.39 \pm 0.28$ & $19.72\pm0.08$ & -- & --  & -- & ZTF \\
59284.36 & $+00.4$ & -- & -- & -- & $20.30\pm0.22$ & -- & ATLAS \\
59288.35 & $+04.3$ & -- & -- & -- & $20.75\pm0.23$ & -- & ATLAS \\
59291.34 & $+07.3$ & -- & $20.10 \pm 0.12$ & -- & -- & -- & ZTF \\
59293.23 & $+09.2$ &$>20.64$ & $20.35 \pm 0.22$ & -- & -- & -- &ZTF \\
59296.26 & $+12.3$ & -- & -- & $20.01\pm0.30$ & --  & -- &ZTF \\
59302.31 & $+18.3$ & -- & -- & $19.87\pm0.22$ & -- & -- &ZTF \\
59304.32 & $+20.3$ & & & & & $20.47\pm0.31$ & ATLAS \\
59306.14 & $+22.1$ & -- & $20.59\pm0.11$ & -- & -- & -- &SEDM \\
59307.16 & $+23.2$ & $>20.83$ & -- & -- & -- & -- &ZTF \\
59308.32 & $+24.3$ & -- & -- & $20.39\pm0.24$ & -- & -- &ZTF \\
59309.24 & $+25.2$ & -- & $20.79\pm0.30$ & -- & -- & -- &ZTF \\
59311.26 & $+27.3$ & -- & -- & $20.57\pm0.27$ & -- & -- &ZTF \\
59315.18 & $+31.2$ & -- & -- & $20.81\pm0.30$ & -- & &ZTF \\
59317.65 & $+33.7$ & -- & $21.27\pm0.21$ & $20.63\pm0.21$ & -- & -- &SEDM \\
59366.88 & $+82.9$ & -- & $22.84 \pm 0.12$ & -- & -- & -- &NOT \\
\hline
\hline
\end{tabular}

\begin{tablenotes} 
\item $a$ : Phase is given in days since \emph{r-}band peak. 
\end{tablenotes}

\end{center}
\end{table*}
\endgroup

\section{Light-curve analysis}
\label{sec:lc}
\begin{figure*}[hbt]
    \centering
    \includegraphics[width=\textwidth]{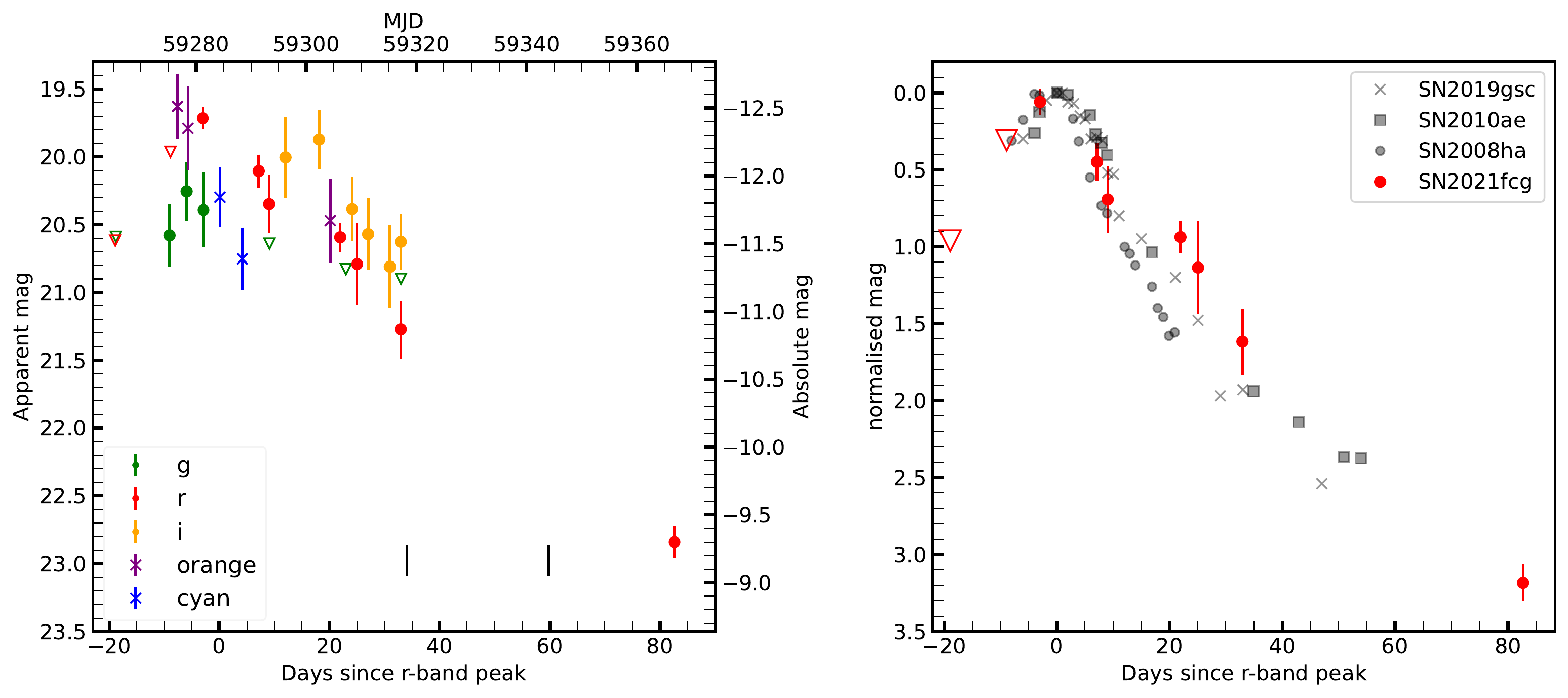}
    \caption{\emph{Left} : \emph{g, r, i, o} and \emph{c} band light curve of SN\,2021fcg. The epochs of our spectroscopic observations are marked with a black vertical line. \emph{Right} : Comparison of the \emph{r}-band photometric evolution to other low luminosity SN Iax (SNe 2008ha, 2019gsc and 2010ae). In the first 12 days post maximum, the evolution of SN\,2021fcg is similar to SNe 2010ae and 2019gsc, but slower than SN\,2008ha. After 12 days, the evolution of SN\,2021fcg slows down, with a possible plateau between 12 and 22 days. This plateau is also seen in the \emph{i}-band measurements (left panel). Data for the comparison objects is taken from \citet{Valenti2009}, \citet{Stritzinger2015} and \citet{Srivastav2020} respectively.}
    \label{fig:lc}
\end{figure*}

Figure \ref{fig:lc} shows the light-curve of SN\,2021fcg. 

We constrain the explosion time between MJD $59265.28<t_{\rm{exp}} < 59275.16$ (based on the latest, deepest non-detection and the first detection). The \emph{g-}band light curve of SN\,2021fcg has only three points that do not show significant evolution and hence samples the supernova around the peak. We report the mean of the three detections as the peak \emph{g-}band magnitude. We also report the mean of the first two \emph{o-}band detections as the peak \emph{o} magnitude, however this is poorly constrained as the latest \emph{o} upper limit prior to first detection is shallower than the first detection. We cannot constrain the \emph{c-}band peak from our two detections. To determine the peak brightness in the \emph{r-}band, we use the light curve of SN\,2019gsc as a template and fit the stretched-scaled template to the observed light curve of SN\,2021fcg. 

The extinction corrected peak apparent magnitudes of SN\,2021fcg are m$_{o}^{\rm{peak}} = $ $19.54\pm 0.28$, m$_{g}^{\rm{peak}} = $ $20.16\pm 0.24$ and m$_{r}^{\rm{peak}} = $ $19.48\pm0.14$ on MJD $59284.0\pm 1.5$ days. This corresponds to M$_{o}^{\rm{peak}} = $ $-12.60\pm0.32$, M$_{g}^{\rm{peak}} = $ $-11.66\pm0.42$ and M$_{r}^{\rm{peak}} = -12.66 \pm 0.20$ mag. 
 
This makes SN\,2021fcg the lowest luminosity SN discovered to date. The faintest previously known SN-Iax are SN\,2019gsc ($M_{\rm{g}}^{\rm{peak}} = -13.58 \pm 0.15$, $M_{\rm{r}}^{\rm{peak}} = -14.28 \pm 0.15$, \citealt{Tomasella2020,Srivastav2020}), SN\,2010ae ($M_{\rm{g}}^{\rm{peak}} = -14.2 \pm 0.5$, $M_{\rm{r}}^{\rm{peak}} = -14.6 \pm 0.5$, \citealt{Stritzinger2014}) and SN\,2008ha ($M_{\rm{g}}^{\rm{peak}} = -13.89 \pm 0.14$, $M_{\rm{r}}^{\rm{peak}} = -14.25 \pm 0.14$, \citealt{Valenti2009}).  SN\,2021fcg is more than a magnitude fainter than these SNe (see Figure \ref{fig:formation_channels}). 

We note that extinction from the host galaxy can increase our estimate of the peak brightness. 
In the absence of any host extinction indicators (Sec. 2.2), we use the $(g-r)$ color of SN\,2021fcg to examine the effect of host extinction on our measurements. The $(g-r)$ color of SN\,2021fcg at the \emph{r}-band peak corrected for Galactic extinction is $0.60\pm0.29$ mag. The corresponding value for SN\,2008ha is 0.57$\pm$0.03 mag, SN\,2019gsc is 0.42$\pm$0.12 mag and SN\,2010ae is 0.42$\pm$0.04 mag \citep{Srivastav2020,Foley2009,Stritzinger2014}. SN\,2008ha and SN\,2019gsc had no significant host extinction, while E$(B-V)_{\rm{host}} = 0.3$ is the most appropriate value for SN\,2010ae \citep{Srivastav2020, Stritzinger2014}. If SN\,2021fcg has similar peak colors as the other low luminosity SNe, the host extinction is E$(B-V)_{\rm{host}} \approx 0.2^{+0.3}_{-0.2}$. Using E$(B-V)_{\rm{host}} = 0.2$ gives M$_{r,\rm{peak}} = $ $-13.16\pm0.20$. Even with the E$(B-V)=0.5$, SN\,2021fcg has M$_{r,\rm{peak}} = $ $-13.78\pm0.20$ and is still among the lowest luminosity thermonuclear supernova discovered to date. However, no relation between SN Iax peak colors has been established, so these extinction estimates are representative at best.

The right panel of Figure \ref{fig:lc} shows the \emph{r}-band photometric evolution of SN\,2021fcg compared to other low luminosity SN Iax. The overall evolution of SN\,2021 fcg is broadly consistent, albeit slightly slower than the other three SNe. However, the slow apparent evolution might be an effect of the epoch of peak brightness not being constrained accurately. We note that 
the \emph{r} and \emph{i}-band light curve flattened between +9 days and +22 days, a behaviour that is not seen for the other three SNe (Figure \ref{fig:lc}). From +25 d to +80 d, the \emph{r}-band light curve declined at a rate of $\approx$ 0.04 mag day$^{-1}$. 

\begin{figure*}[hbt]
    \centering
    \includegraphics[width=0.6\textwidth]{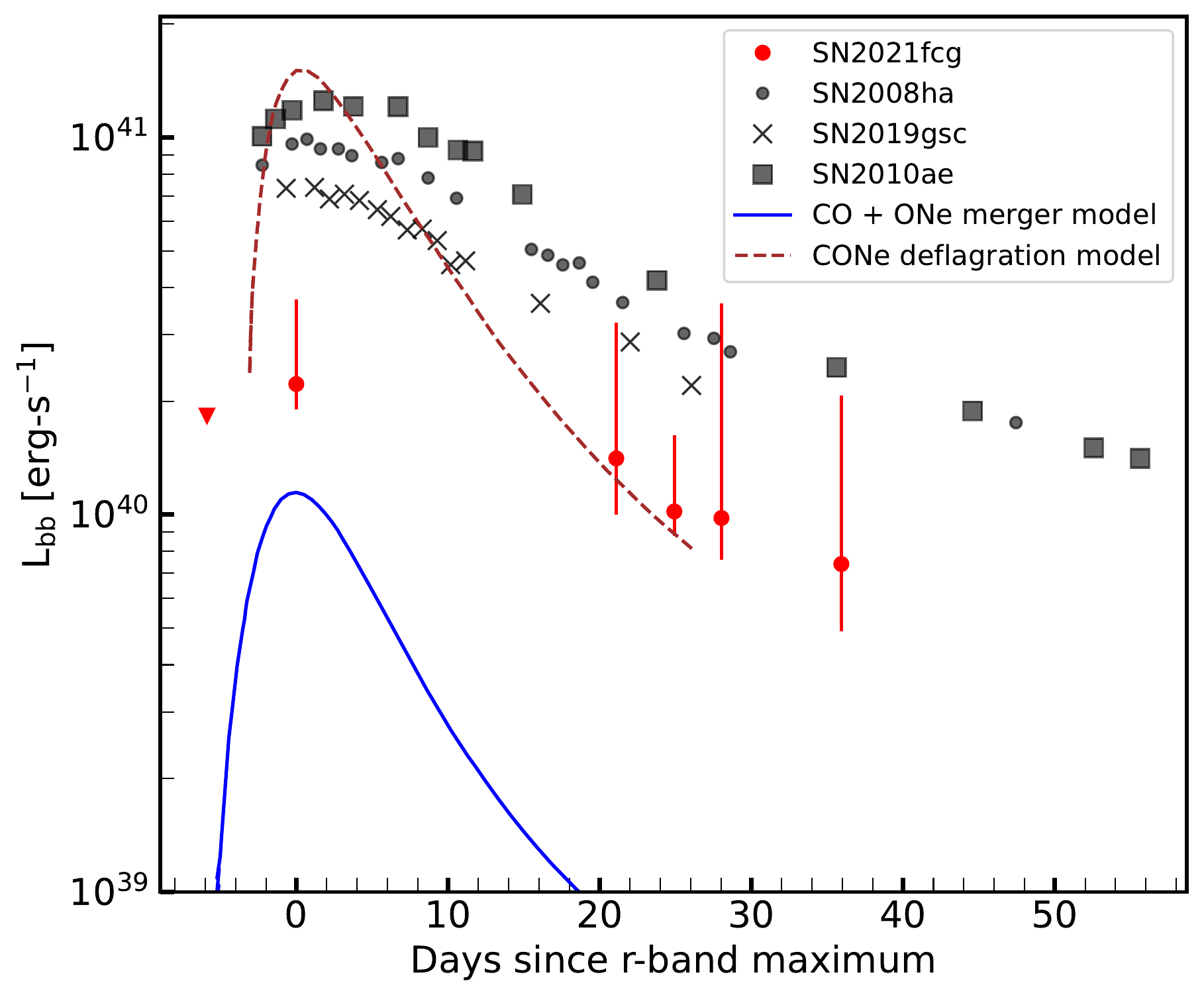}
    \caption{Bolometric luminosity evolution (derived from blackbody fitting) of SN\,2021fcg compared with the blackbody luminosities of SNe 2008ha, 2010ae and 2019gsc \citep{Srivastav2020}. The peak luminosity of SN\,2021fcg is smaller than the other low-luminosity SN Iax by a factor of $\sim 3$. Also plotted are the theoretical bolometric luminosities of explosions from deflagration of a near-M$_{\rm{ch}}$ white dwarf (dashed-brown line, \citealt{Kromer2015}) and the merger of a 1.1 M$_{\odot}$ CO and a 1.2 M$_{\odot}$ ONe white dwarf (solid blue line, \citealt{Kashyap2018}). }
    \label{fig:lum_bb}
\end{figure*}

\subsection{Bolometric luminosity}
We fit the photometric measurements with a blackbody function to derive the bolometric luminosity of the supernova. As our lightcurve sampling is sparse, we do not have contemporaneous multiband observations. We interpolate between the \emph{r} and \emph{i}-band detections using a Gaussian process with a radial basis function (RBF) kernel to generate synthetic measurements wherever necessary. The Gaussian process model was implemented using \texttt{scikit-learn}\footnote{https://scikit-learn.org/stable/modules/gaussian\_process.html} We then fit a blackbody function to these measurements with a Markov Chain Monte Carlo (MCMC) analysis using the \texttt{python} package \texttt{emcee} \citep{Foreman2013} to derive the effective temperatures, photospheric radii and bolometric luminosities. We note that the systematic uncertainties on our estimates are large as they are derived using data for only two filters. Additionally, we do not have any ultraviolet (UV) or near-infrared (NIR) photometric coverage. For SN2019gsc, \citet{Srivastav2020} estimated that the UV and NIR contribution increases the optical blackbody luminosity by a factor of $\approx1.5$. Assuming a similar contribution for SN\,2021fcg, we add a 50\% systematic uncertainty to our luminosity estimates. We note that the late time spectra do not resemble a blackbody. However, we use simple blackbody estimates as the data available is limited.

Figure \ref{fig:lum_bb} shows the evolution of the bolometric luminosity of SN\,2021fcg compared to SN\,2008ha, SN\,2010ae and SN\,2019gsc (taken from \citealt{Srivastav2020}). The peak bolometric luminosity is 2.5$^{+1.5}_{-0.3} \times 10^{40}$ erg s$^{-1}$ which is $\sim$ 3 times lower than SN\,2019gsc. We model the bolometric luminosity evolution using the relations from \citet{Arnett1982} formulated as in \citet{Valenti2008}. This model assumes that the light curve is powered by radioactive decay of $^{56}$Ni and $^{56}$Co in the ejecta. The ejecta are assumed to be spherically symmetric, homologously expanding and have constant opacity (see \citealt{Valenti2008} for additional details). Under these assumptions, the evolution of the bolometric luminosity can be described using three parameters  --  the total nickel mass M$_{\rm{Ni}}$, the lightcurve timescale ($\tau_{\rm{M}}$) and the explosion time ($\tau_{\rm{exp}}$). We use \texttt{emcee} to estimate the best-fit parameters for SN\,2021fcg. We derive a nickel mass of M$_{\rm{Ni}} = 0.8^{+0.4}_{-0.5} \times 10^{-3}$ M$_{\odot}$. This is lower than the nickel mass in SN\,2019gsc ($1.4-2.4 \times 10^{-3}$ M$_{\odot}$), SN\,2008ha ($3\times10^{-3}$ M$_{\odot}$) and SN\,2010ae ($3-4\times10^{-3}$M$_{\odot}$, derived in \citealt{Srivastav2020}).

The total ejecta mass (M$_{\rm{ej}}$) can be derived using the relation M$_{\rm{ej}} = \frac{1}{2}\tau_{M}^{2}\frac{\beta c v_{\rm{peak}}}{k_{\rm{opt}}}$  \citep{Valenti2008}, where $\beta=13.8$, \emph{c} is the speed of light, $v_{\rm{peak}}$ is the peak photospheric velocity and $k_{\rm{opt}}$ is the net opacity. We cannot estimate $v_{\rm{peak}}$ for SN\,2021fcg as we do not have spectroscopic coverage near maximum light. We assume $v_{\rm{peak}} = 3500$ km s$^{-1}$ and $k_{\rm{opt}} = 0.1$ cm$^{2}$g$^{-1}$ similar to SN\,2019gsc \citep{Srivastav2020}, we derive M$_{\rm{ej}} = 0.05 - 0.4$ M$_{\odot}$.

Finally, we note that for E(B-V) = 0.2 and 0.5 respectively, the peak luminosity of SN\,2021fcg is 4.3$_{-0.6}^{+2.0} \times 10^{40}$erg s$^{-1}$ and $1.1_{-0.3}^{+3.0} \times 10^{41}$ erg s$^{-1}$ respectively.

\section{Spectroscopic evolution}
\label{sec:spectroscopy}

\begin{figure*}[hbt]
    \centering
    \includegraphics[width=\textwidth]{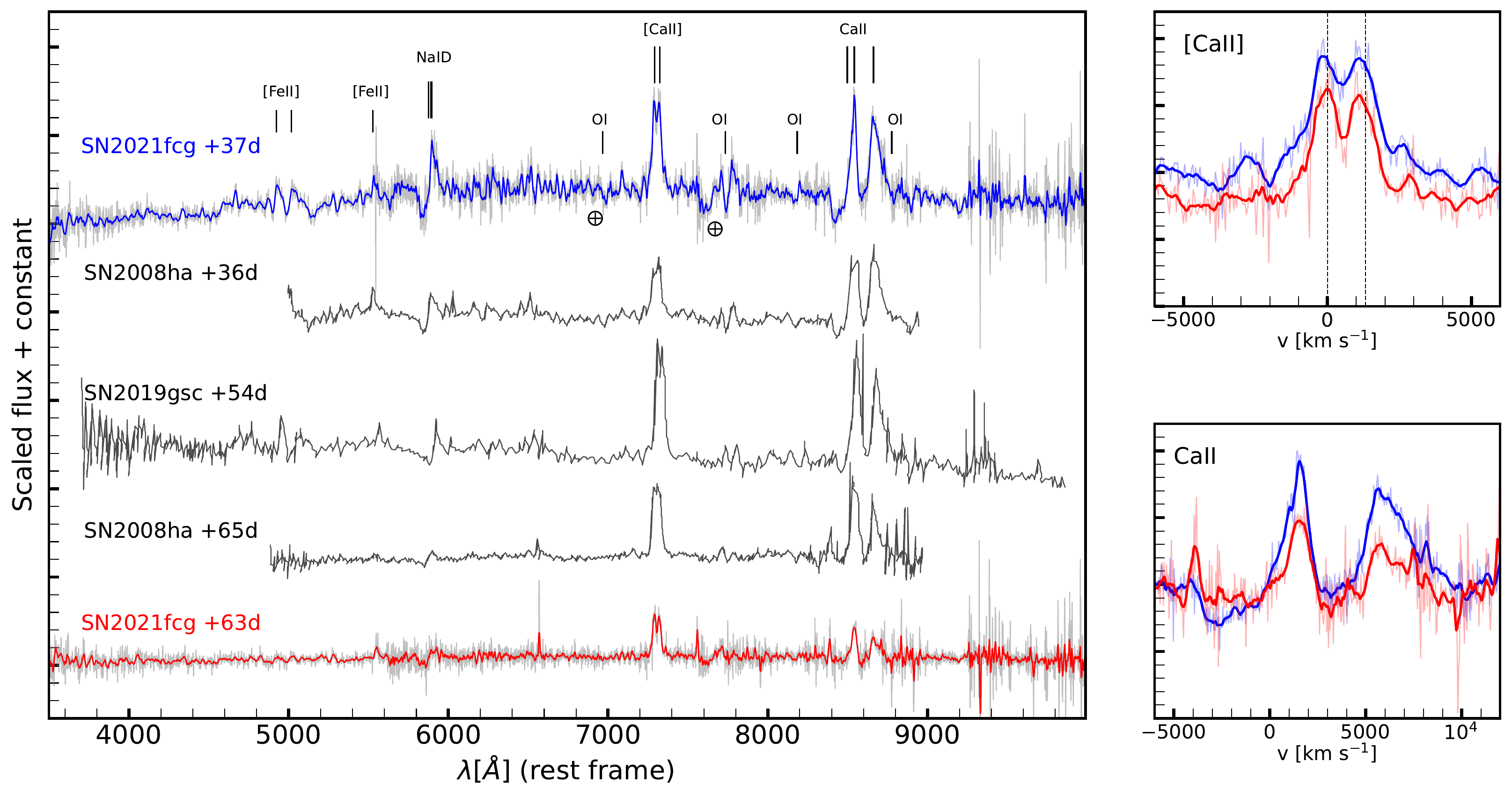}
    \caption{\emph{Left} : Late-time optical spectra of SN\,2021fcg at +37d (blue) and +63d (red). Also plotted are similar phase spectra of SN\,2008ha and SN\,2019gsc (gray, taken from \citealt{Valenti2009, Tomasella2020}). The spectra of SN\,2021fcg closely resemble the spectra of the other low-luminosity SN Iax. \emph{Right}: Zoom-in of the [CaII] \emph{(top)} and CaII \emph{(bottom)} emission lines. These are the strongest features in the spectra (37d : blue, 63d : red ). In both spectra, the peak wavelengths of [CaII] emission are blueshifted from the rest wavelengths (black dashed lines in top panel) by $\approx 100$ km-s$^{-1}.$ }
    \label{fig:spectra}
\end{figure*}

Figure \ref{fig:spectra} shows the +37d and +63d (rest frame phase from peak) optical spectra of SN\,2021fcg. Both spectra show characteristics of late-time SN Iax spectra and closely resemble similar phase spectra of SN\,2008ha and SN\,2019gsc \citep{Valenti2009,Tomasella2020}. We do not detect any hydrogen, helium or oxygen lines in our spectra that could be indicative of a nova. In the nebular phases, the calcium lines in SN\,2021fcg have velocities of $\approx1500$ km s$^{-1}$ -- much smaller than those measured for Ca-rich SNe ($\approx$ 5000 km s$^{-1}$ \citealt{De2020a}). Overall, our spectra strongly indicate that SN\,2021fcg is a low luminosity SN Iax.


The +37d spectrum shows a slightly reddened continuum with several emission lines. The strongest emission features are the [\ion{Ca}{2}] doublet and \ion{Ca}{2} NIR triplet. Both of the [\ion{Ca}{2}] doublet lines have a FWHM of $\approx$1300 km s$^{-1}$ (measured using Gaussian fits). We also detect a possible absorption feature for these lines at a velocity of $-3500$ km s$^{-1}$ (see Figure \ref{fig:spectra}). For the \ion{Ca}{2} NIR triplet, the 8498 and 8548\AA~ lines are blended together, with a FWHM of 1330 km s$^{-1}$. These lines also show a P-cygni profile with the emission maximum at $\sim 1500$ km s$^{-1}$ and the absorption minimum at $-2500$ km s$^{-1}$. However, the absorption is likely affected by blending from other absorption lines. For the 8662~\AA~ line, only an emission component is detected with a FWHM of 2620 km s$^{-1}$. Similar features are also seen in the spectra of SN\,2019gsc and SN\,2008ha. The spectrum also shows strong Na I D emission with a P-cygni profile, although this absorption is also affected by blending. In addition, we detect several [\ion{Fe}{2}] emission lines. We note that the \ion{O}{1} emission is weak and no Ni or He features are detected.

The +63d spectrum is similar, with a very weak continuum. The [\ion{Ca}{2}] doublet is the strongest emission feature with a FWHM of $\approx$ 1000 km s$^{-1}$. The \ion{Ca}{2} NIR triplet is also detected in emission, with a FWHM of $\approx$1050 km s$^{-1}$ (8498 and 8548 \AA) and 1850 km s$^{-1}$ (8662 \AA). Compared to the +37 d spectrum, the \ion{Na}{1} emission is very weak. 

These spectral features and FWHM velocities are similar to those seen in other low-luminosity SN Iax. 
We also note that in both spectra, the peak wavelengths of the [\ion{Ca}{2}] emission lines are blueshifted by $\approx 100$ km s$^{-1}$ with respect to the rest wavelength. 

\section{Discussion}
\label{sec:discussion}
The extremely low luminosity of SN\,2021fcg makes it a remarkable member of the class of thermonuclear supernovae. In Figure \ref{fig:formation_channels} (left panel), we compare the peak absolute magnitude and decline rate of SN\,2021fcg to other thermonuclear supernovae. While the peak absolute magnitude of SN\,2021fcg is $\approx$1.5 mag fainter than all other supernovae, its decline rate is not extreme ($\Delta m_{15,r} = 0.7\pm0.3$ mag), and is similar to the other three low luminosity SN Iax. 


\subsection{What is the progenitor of SN\,2021fcg?}
\label{sec:origin}
The exact origin of SN Iax is debated, however it is generally accepted that they are associated with thermonuclear explosions of white dwarfs \citep{Jha2017}. In particular, models involving the deflagration of a near Chandrasekhar (M$_{\rm{ch}}$) accreting white dwarf that leaves behind a bound remnant have been able to reproduce some observed properties of SN Iax. Failed deflagrations of a near M$_{\rm{ch}}$ CO white dwarf successfully explain several features of the luminous SN Iax \citep{Jordan2012,Kromer2013,Fink2014}. However, this model cannot explain the lower luminosity 08ha-like explosions $-$ the faintest supernova from the CO deflagration model has M$_{V} = -16.8$ mag \citep{Fink2014}. 

\citet{Kromer2015} proposed that the low luminosity SN Iax could be deflagrations of hybrid CONe white dwarfs \citep{Chen2014, Denissenkov2015}. The presence of the ONe layer can quench the burning to suppress the amount of nickel produced. They simulated the deflagration of a 1.4 M$_{\odot}$ CONe white dwarf with five ignition cores and found that this results in a low luminosity ($M_{\rm{V}} \approx -14.2$) transient roughly consistent with SN 08ha, 10ae and 19gsc. However, they derive a total ejecta mass of 0.014 M$_{\odot}$ -- an order of magnitude smaller than SN 08ha, 10ae and 19gsc. Consequently, the transient in their simulations evolves faster than these three SN (see Figure \ref{fig:lum_bb}).


SN\,2021fcg is fainter than the CONe deflagration model by $\approx 1.5$ mag. In order to explain the luminosity and timescale of SN\,2021fcg as the outcome of a white dwarf deflagration, the nickel ejecta mass has to be lower by a factor of $\sim 3$ and the total ejecta mass has to be at least 10 times larger than the value obtained by \citet{Kromer2015}. It remains to be seen if lower number of ignition cores or a larger ONe mass than the one used by \citet{Kromer2015} can reduce the nickel yield enough to explain the low luminosity of SN\,2021fcg. However, a lower total ejecta mass will make the transient even faster evolving and inconsistent with the slow decline rate of SN\,2021fcg. SN\,2021fcg is thus a challenge to the existing picture of hybrid white dwarf deflagration and warrants further investigation of this channel. 


\begin{figure*}[hbt]
    \centering
    \includegraphics[width=\textwidth]{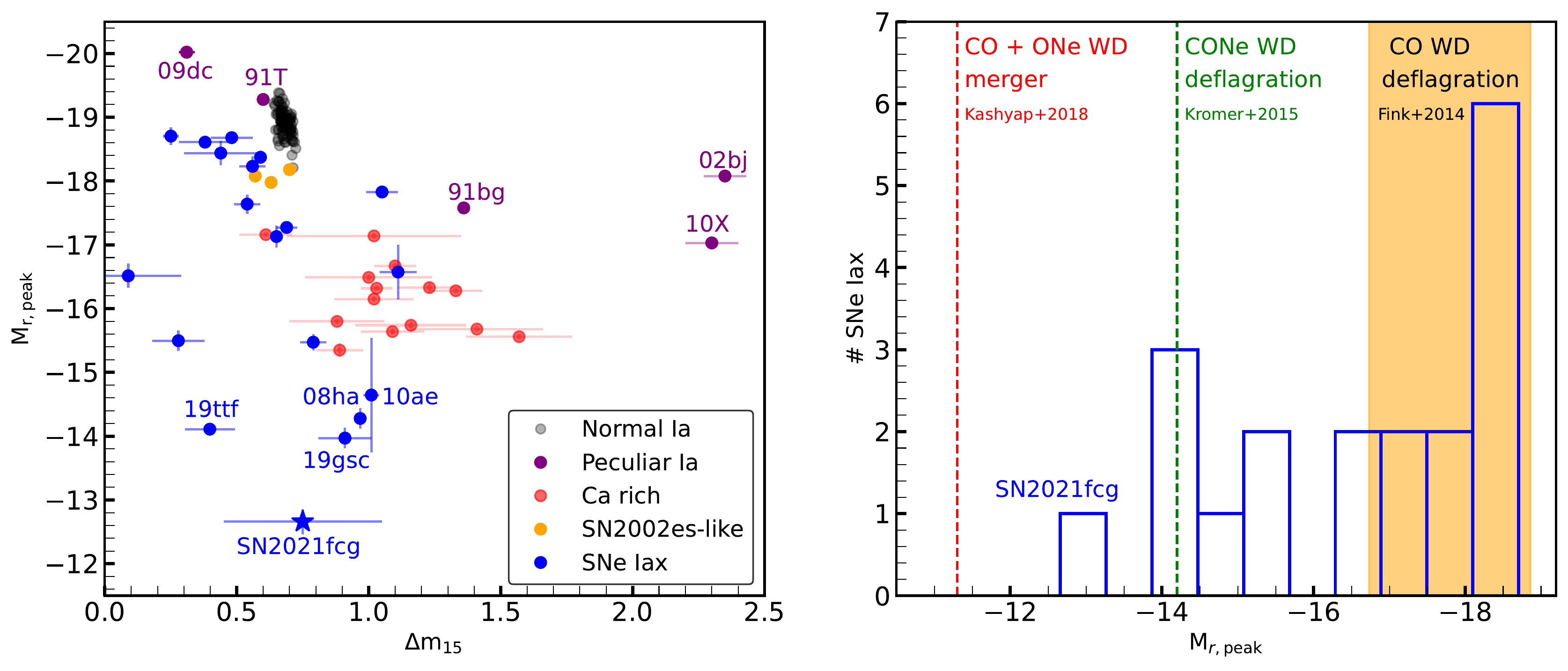}
    \caption{\emph{Left}: Peak \emph{r-}band absolute magnitudes against \emph{r-}band decline rates ($\Delta m_{15}$) of thermonuclear supernovae. SN\,2021fcg (blue star) is the least luminous thermonuclear SN discovered to date. We also plot the values for normal SNe Ia (black circles, \citealt{Yao2019}), the peculiar SNe Ia 1991bg \citep{Filippenko1992}, 1991T \citep{Lira1998}, 2009dc \citep{Taubenberger2011}, 2010x \citep{Kasliwal2010} and 2002bj \citep{Poznanski2010} (purple circles), calcium-rich gap transients (red circles, \citealt{De2020a}), 02es-like transients (orange circles, \citealt{White2015}), SNe Iax (blue circles, \citealt{Magee2016,Foley2013,Srivastav2020}). We use the Tully Fisher distance modulus for SN\,2021fcg, while we use H$_{0} = 73$ km s$^{-1}$ Mpc$^{-1}$ for all other SNe. \\ \emph{Right:} A histogram of peak SN Iax absolute magnitudes together with the predictions of existing progenitor models. The bright (M$_{r} < -16.8$ mag) SN Iax are consistent with simulations of deflagrations of near M$_{\rm{ch}}$ CO white dwarfs (orange region, \citealt{Fink2014}). The lower luminosity 08ha-like SNe are consistent with the simulation of a deflagration of hybrid CONe white dwarf (green line, \citealt{Kromer2015}). SN\,2021fcg is even fainter, and it is unclear if it could be explained by different initial conditions of the CONe WD deflagration. A promising alternative is the double degenerate model involving the merger of a 1.1 M$_{\odot}$ CO and 1.2 M$_{\odot}$ ONe white dwarf. A more massive ONe white dwarf could give rise to more luminous transients such as SN\,2021fcg. }
    \label{fig:formation_channels}
\end{figure*}

An alternative picture for formation of low luminosity SNe Iax is the double-degenerate scenario. \citet{Kashyap2018} showed that the merger of a CO and ONe white dwarf yields a failed detonation of the ONe core producing small amounts of ejecta. This gives rise to a very faint, rapidly evolving transient. They modeled the merger of a 1.1 M$_{\odot}$ CO and a 1.2 M$_{\odot}$ ONe white dwarf and derive a very low nickel yield  ($5.7\times10^{-4}$ M$_{\odot}$). The resulting transient has peak M$_{V} = -11.3$, an ejecta velocity of $\sim 4000$ km s$^{-1}$ and a total ejecta mass of $\sim 0.08$ M$_{\odot}$.  

The observed luminosity of SN\,2021fcg is $\sim 3$ times brighter than that predicted by \citet{Kashyap2018} (see Figure \ref{fig:lum_bb}), while the nickel and total ejecta masses are roughly consistent with their estimates. The double-degenerate channel is thus a promising model to explain SN\,2021fcg, as a merger involving a more massive ONe white dwarf could give a brighter and longer lived transient than the one modelled in \citet{Kashyap2018}. However, the requirement of a more massive ONe white dwarf may decrease the expected rate of such explosions. Additional studies of this model are necessary to test its viability as the progenitor of SN\,2021fcg. In this channel, the remnant is a kicked super-Chandrasekhar star with an ONe core embedded in the nebulosity of the SN ejecta. Recently, \citet{Oskinova2020} identified a candidate super-M$_{\rm{ch}}$ remnant at the center of the nebula IRAS00500+6713. They posit that this source is the remnant of a SN Iax resulting from an ONe and CO white dwarf merger. If SN\,2021fcg is the result of a white dwarf merger, the remnant would be a similar super-M$_{\rm{ch}}$ star that will eventually end its life as an electron-capture SN.

In any of these scenarios, the surviving bound remnant can drive winds from its surface through delayed radioactive decay \citep{Shen2017}. These winds can be the dominant source of luminosity at late times ($\approx$ 1 year post explosion) lasting for at least a decade. 
Recent \emph{Hubble Space Telescope} (HST) observations of SN\,2012Z revealed an excess in its late-time luminosity that could be a result of remnant-driven winds \citep{McCully2021}. Late time HST observations of SN\,2021fcg will be valuable in probing the nature of its bound remnant. 

We summarize these possible formation scenarios in the right panel of Figure \ref{fig:formation_channels}. Additional scenarios that do not invoke a white dwarf have also been proposed to explain other low luminosity SNe Iax. For example, \cite{Valenti2009} note that SN\,2008ha has similarities with models of low luminosity core-collapse SN such as a fallback massive star SN \citep{Moriya2010} or an ultra-stripped electron-capture SN \citep{Pumo2009}. NIR spectra can help distinguish between a thermonuclear and core-collapse origin for these supernovae \citep{Stritzinger2015}.

\section{Conclusions}
\label{sec:conclusions}
We have presented optical photometry and late-time optical spectroscopy for SN\,2021fcg -- the lowest luminosity SN-like transient discovered to date. The photometric and spectroscopic evolution of SN\,2021fcg closely resembles low luminosity SN Iax such as SN\,2008ha, SN\,2010ae and SN\,2019gsc. SN\,2021fcg has M$_{r} = $ $-12.66\pm0.20$ mag and is the faintest of the faint SN Iax, fainter than the other members by more than a magnitude. The lower luminosity, lower nickel ejecta mass and slightly slower photometric evolution of SN\,2021fcg represent a challenge to theoretical models of low luminosity SN Iax. Existing hybrid CONe white dwarfs deflagration models are overluminous by a factor of $\sim 3$. A double degenerate scenario in which the SN is an outcome of a CO and ONe white dwarf merger is a promising model. Formation channels that involve a core-collapse origin are also plausible, but unlikely.

Additional observations of SNe Iax with extremely low luminosities (M$\approx -12.5$ mag) are required to identify the explosion mechanisms of these mysterious transients. ZTF can detect these SNe to a distance of $\approx 40$ Mpc.  The Vera Rubin Observatory (VRO, \citealt{Ivezic2019}) will significantly increase the discovery distance to $\approx 275$ Mpc. However, given the fast evolution of these transients, rapid followup observations will be required to derive useful insights about them. An experiment similar to the Census of the Local Universe \citep{De2020a} that keeps track of VRO transients in catalogued galaxies will be instrumental in discovering such low luminosity SNe.


\section{Acknowledgements}
We thank the anonymous reviewer for comments that helped improve the paper.
We thank Harsh Kumar, Varun Bhalerao and G.C. Anupama for photometric observations with the GROWTH-India telescope \footnote{https://sites.google.com/view/growthindia/}. 
This paper is based on observations obtained with the Samuel Oschin Telescope 48-inch and the 60-inch Telescope at the Palomar Observatory as part of the Zwicky Transient Facility project. ZTF is supported by the National Science Foundation under Grant No. AST-2034437 and a collaboration including Caltech, IPAC, the Weizmann Institute for Science, the Oskar Klein Center at Stockholm University, the University of Maryland, Deutsches Elektronen-Synchrotron and Humboldt University, the TANGO Consortium of Taiwan, the University of Wisconsin at Milwaukee, Trinity College Dublin, Lawrence Livermore National Laboratories, and IN2P3, France. Operations are conducted by COO, IPAC, and UW. SED Machine is based upon work supported by the National Science Foundation under Grant No. 1106171. This work is also based on observations made with the Nordic Optical Telescope, owned in collaboration by the University of Turku and Aarhus University, and operated jointly by Aarhus University, the University of Turku and the University of Oslo, representing Denmark, Finland and Norway, the University of Iceland and Stockholm University at the Observatorio del Roque de los Muchachos, La Palma, Spain, of the Instituto de Astrofisica de Canarias.  SED Machine is based upon work supported by the National Science Foundation under Grant No. 1106171. The ZTF forced photometry service was funded under the Heising-Simons Foundation grant 12540303 (PI: Graham). KM is funded by the EU H2020 ERC grant no. 758638. SS and ECK acknowledge support from the G.R.E.A.T research environment, funded by {\em Vetenskapsr\aa det},  the Swedish Research Council, project number 2016-06012. ECK acknowledges support from The Wenner-Gren Foundations. MR has received funding from the European Research Council (ERC) under the European Union's Horizon 2020 research and innovation programme (grant agreement 759194 - USNAC)


\bibliography{myreferences}
\bibliographystyle{apj}

\label{lastpage}
\end{document}